\documentclass[twocolumn,pre]{revtex4}
\usepackage{amssymb,bbm,bm}
\usepackage{graphicx}
\usepackage{bm}
\usepackage{amsmath}
\usepackage[usenames]{color}
\usepackage{epsfig}
\usepackage{hyperref}
\usepackage{subfigure}
\usepackage[sans]{dsfont}
\hypersetup{colorlinks=true, citecolor=blue,
linkcolor=blue,urlcolor=blue }

\usepackage{amsmath,amsfonts,amssymb,amsthm,bm}
\usepackage{float}
\usepackage{hyperref}
\usepackage{graphicx}
\usepackage{subfigure}

\begin{document}

\title{Griffiths singularity in quasi-one dimensional restricted $\pm J$ Ising spin glass}
\author{Fateme Izadi}
\author{Reza Sepehrinia}\email{sepehrinia@ut.ac.ir}
\affiliation{Department of Physics, University of Tehran, Tehran 14395-547, Iran}

\begin{abstract}
We obtain the exact ground state energy of the quasi-one dimensional restricted $\pm J$ Ising spin glass in a uniform magnetic field using the transfer matrix method. Magnetic field dependence allows us to derive the magnetization as a function of concentration and magnetic field. It turns out that, in the limit of zero field, the magnetization tends to a nonzero value with a singular dependence on the magnetic field. We derive the explicit form of the singularity in thermodynamic quantities such as energy $E\simeq E_0+m_0 h + E_1e^{-h_0/h}$, which is an essential singularity known as Griffiths singularity. We confirm our analytical results using the numerical approach based on iterative equations for energy.

\end{abstract}
\maketitle

\section{Introduction}

Spin glasses as the prototype of the glassy systems are a subject of interest for many disciplines in science. From the fundamental point of view, the physics of these systems is not fully understood. The question of spin glass being a new magnetic phase of matter is somewhat open \cite{charbonneau2023spin}. Besides the fundamental importance spin glass concepts, ideas, and mathematical tools were applied to problems in neural networks, combinatorial optimization, biological evolution, protein dynamics and folding, computer science, mathematics, and the social sciences \cite{stein2013spin,hertz1991introduction,mezard2009information,nishimori2001statistical,huang2022statistical}.

Most of our theoretical knowledge on spin glasses is based on the mean field theory or infinite-range Sherington Kirkpatrick model \cite{sherrington1975solvable} whose thermodynamic behavior is identical to the short-range model in infinite dimensions. The theory of the short-range model in finite dimensions mainly relies on numerical simulations except in one dimension where analytical proof exists. The current consensus is that there is no transition in dimensions less than three at a finite temperature. There is however evidence of true long-range spin glass order at zero temperature in two dimensions \cite{jorg2006strong,roma2010ground,hartmann2023metastate,middleton1999numerical}.

Therefore the exact analytical results in low dimensional systems can be helpful in understanding the physics of the spin glass phase. Especially the one-dimensional and quasi-one dimensional models which are feasible for analytical treatment. The ground state energy and magnetization of linear chain with $\pm J$ \cite{vilenkin1978random,derrida1978simple,ardebili2016ground} and continuous \cite{chen1982low,gardner1985zero} distribution of couplings in the presence of a uniform magnetic field has been investigated. For linear chain, the magnetization vanishes at zero field limit with a power-law dependence on the magnetic field for both discrete and continuous coupling distribution. For strips of a few coupled chains, it is also possible to calculate the ground state energy analytically. Three coupled chains with periodic boundary condition in the transverse direction and $\pm J$ distribution has been investigated by Derrida et al. \cite{derrida1978simple}. The linear triangular chain has been discussed in Refs. \cite{morita1980spin,fechner1986ground}. The method of Ref. \cite{derrida1978simple} is generalized to calculate the ground state energy of strips with larger widths in Ref. \cite{kadowaki1996exact}.

In order to obtain the magnetization one needs the dependence of the energy on the magnetic field. Including the magnetic field makes the problem more complicated. For the cases in which the enumeration of the spin configurations is possible, the inclusion of the magnetic field can be done more easily. One such example which is investigated in Ref. \cite{timonin2003thermodynamics}, is the Ising ladder with randomness only in transverse links with certain constraints on couplings. But in general, the enumeration of clusters that flip by increasing the magnetic field is not a simple problem. The transfer matrix method \cite{derrida1978simple} is suitable as it allows the calculations without enumeration of spin configurations. This method is implemented numerically and magnetic field dependence of energy and magnetization has been investigated for strips of various widths by one of authors and a collaborator\cite{sepehrinia2018ground}.

In this paper, we study a simplified version of $\pm J$ Ising spin glass on ladder structure in the presence of the magnetic field. We use the transfer matrix method to obtain the exact analytical expression for zero temperature energy and magnetization. In contrast with the continuous distribution  \cite{sepehrinia2018ground}, where the magnetization tends to zero in zero field, in this case, we find nonzero magnetization with essential singularity in magnetic field dependence in zero field limit. This kind of singularity was first discovered by Griffiths \cite{griffiths1969nonanalytic} in diluted ferromagnets. Other studies suggest that such a singularity exists also in spin glasses \cite{randeria1985low,matsuda2008distribution}.
\section{Model And Transfer matrix Formalism}\label{sec2}

The Hamiltonian of the $\pm J$ Ising spin glass on the ladder in the uniform magnetic field is given by
\begin{eqnarray} \label{hamiltonian}
\mathcal{H} &=& - \sum_{i=1}^{L-1} (J_{i,1}^{h} \sigma_{i,1} \sigma_{i+1,1} +J_{i,2}^{h} \sigma_{i,2} \sigma_{i+1,2}  \nonumber \\
&& \hspace{1.5cm} + J_{i}^{v} \sigma_{i,1} \sigma_{i,2} ) - h \sum_{i=1}^{L} ( \sigma_{i,1} + \sigma_{i,2}  ),
\end{eqnarray}
where $\sigma_{ij}=\pm 1$, couplings $J^{h},J^{v}$ take $\pm J$ randomly and, $h$ is a uniform magnetic field. As is well-known, the partition function of this quasi-1D system can be expressed in terms of the trace of the product of the transfer matrices $ Z= \text{Tr}(\mathcal{M}_L) $ where $\mathcal{M}_L=\prod_{i=1}^{L-1} M_i$ and $L$ is the length of the ladder. By increasing the length by one unit, three couplings will be added to the ladder. There are $2^3=8$ possibilities for these three couplings and each of them can be represented by a transfer matrix $M_i$. Therefore there will be $8$ matrices and the total transfer matrix of the ladder will be a product of a random sequence of these matrices.

\begin{figure}[t]
\begin{center}
\includegraphics[scale=0.5]{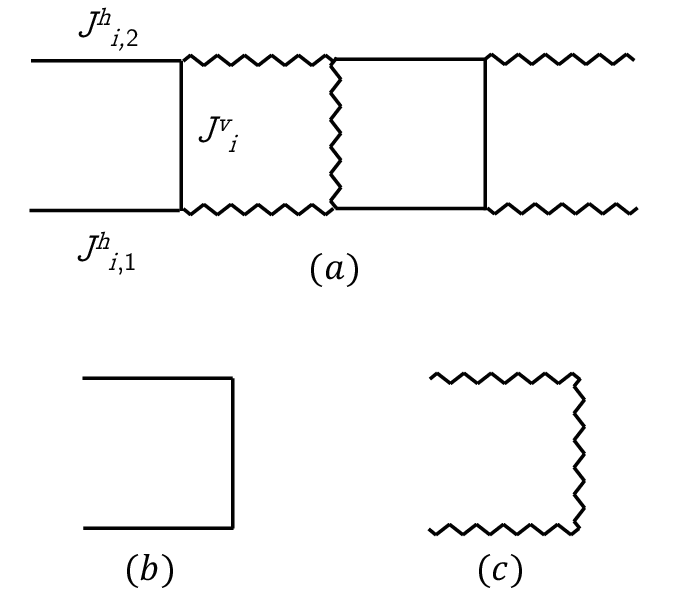}
\end{center}
\caption{(a) The Ising ladder with random $\pm J$ interactions. The wavy lines indicate antiferromagnetic interactions ($-J$). Unit cells with (b) all ferromagnetic couplings (c) antiferromagnetic couplings}
\label{ladder}
\end{figure}

To make this model tractable we simplify it by keeping only two of the matrices, one with probability $x$ and the other with probability $1-x$. We call the resulting model, restricted $\pm J$ Ising spin glass. The schematic diagram of the system is shown in Fig. \ref{ladder} (a), which is composed of two units shown in Fig. \ref{ladder}(b) and Fig. \ref{ladder}(c). The corresponding transfer matrices of these units are given by
\begin{eqnarray}\label{M1M2}
M_1 &=&
\begin{pmatrix}
z^{3 + 2\alpha} & z^{1 + 2\alpha} & z^{1 + 2\alpha} & z^{-1 + 2\alpha} \\
z^{-1} & z & z^{-3} & z^{-1} \\
z^{-1} & z^{-3} & z & z^{-1} \\
z^{-1 - 2\alpha} & z^{1 - 2\alpha} & z^{1 - 2\alpha} & z^{3 - 2\alpha}
\end{pmatrix}, \nonumber\\ \nonumber\\
M_2 &=&
\begin{pmatrix}
z^{-3 + 2\alpha} & z^{-1 + 2\alpha} & z^{-1 + 2\alpha} & z^{1 + 2\alpha} \\
z & z^{-1} & z^{3} & z \\
z & z^{3} & z^{-1} & z \\
z^{1 - 2\alpha} & z^{-1 - 2\alpha} & z^{-1 - 2\alpha} & z^{-3 - 2\alpha}
\end{pmatrix},\nonumber\\
\end{eqnarray}
where $ z = \exp(\beta J) $ and $ \alpha = h/J $.

In selecting two matrices we have noted to preserve the main ingredients of the original model. As the resulting model contains a finite concentration of frustrated plaquettes and the enumeration of the overturned spins is not simple even for the zero magnetic field, therefore we expect to capture the physics of the $\pm J$ spin glass.

Since these matrices have positive elements, the trace in the partition function can be replaced with any element of $ \mathcal{M}_L$ \cite{derrida1978simple}. By considering the evolution of each element of $ \mathcal{M}_L $ under successive matrix products
\begin{equation}
\mathcal{M}_{L+1} = \mathcal{M}_L M_L,
\label{TM}
\end{equation}
the free energy per spin can be obtained directly.

\section{Zero temperature limit}

At the low-temperature limit, $ z \to \infty $, we can keep only the leading term in $z$ for each element of $ \mathcal{M}_L $ under successive matrix products. The leading terms in the first row of $\mathcal{M}_{L}$ and $\mathcal{M}_{L+1}$ can be considered as \cite{derrida1978simple}
\begin{eqnarray} \label{Matrow}
\mathcal{M}_L  &\simeq&
\begin{pmatrix}
y_1 z^a  & y_2 z^{b} & y_3 z^{c} & y_4 z^{d} \\
\ddots & \  & \  & \vdots \\
\ & \  & \ddots & \vdots \\
\cdots & \cdots & \cdots & \
\end{pmatrix}, \notag\\ \notag\\ \notag\\
\mathcal{M}_{L+1} &\simeq&
\begin{pmatrix}
Y_1 z^A  & Y_2 z^{B} & Y_3 z^{C} & Y_4 z^{D} \\
\ddots & \  & \  & \vdots \\
\ & \  & \ddots & \vdots \\
\cdots & \cdots & \cdots & \
\end{pmatrix}, \ \ \notag\\
\end{eqnarray}
Using Eq. (\ref{TM}), one can obtain $ A, B, C, D $ knowing $ a, b, c, d $ and $ M_L $. The equations of evolution of these exponents are given in Appendix A. The free energy can be expressed in terms of these exponents and their probabilities. It turns out that  $b$ and $c$ remain equal along the ladder because the second and third columns of the transfer matrices $M_1$ and $M_2$ have the same set of values. Therefore $b$ and $c$ remain equal as we multiply $\mathcal{M}_{L}$ with any of the transfer matrices $M_1$ and $M_2$. Also, we can see that  $ -2 \leq a-b \leq 2 $ and  $ -4 \leq a-d \leq 4 $. Using the equations given in Appendix A, by multiplication of $M_1$, $A-B$ always remains less than 2 and by multiplication of $M_2$, it will remain bigger than -2. For example in Eq. (\ref{M1A1}) we can see that $A-B=2$ or in the third set Eq. (\ref{M1A3}) we have $A-B=a-d+2+4\alpha$ and using $-4  \leq a-d \leq  -4\alpha$ we have $A-B \leq 2$ and so on. Similarly, it can be shown that $A-D \leq 4$ for $M_1$ and $A-D \geq -4$ for $M_2$.

The number of possible values of the exponents depends on $\alpha$ and can be very large which makes the calculation tedious. In order to reduce the number of values of the exponents, for $\alpha<1$, we assume $ \alpha = 1/n $ where $n$ only takes positive integer values. To obtain the possible values of the exponents, we start with an arbitrary set of values and then by iteration of equations of evolution (Appendix A), we find a closed set of values. For $\alpha=\frac1n, n=2,3,\dots$ we have the following values for $(a-b,a-d)$
\begin{eqnarray}
\begin{cases}
(2 \alpha n , 2 \alpha i), \ \quad 0 \leq i \leq 2n \rightarrow  \ p_i, \\
(-2 \alpha  , 2 \alpha (i-2)), \ \quad 0 \leq i \leq n-1 \rightarrow \ q_i, \\
(2 \alpha (n-i) , -2 \alpha i), \ \quad 0 \leq i \leq n-1 \rightarrow \ r_i, \\
(-2 \alpha (i+1) , -2 \alpha (i+2)), \ \quad  0 \leq i \leq n-1 \rightarrow \ s_i, \\
(2 \alpha  , 0), \ \rightarrow \ t_1, \\
(2 \alpha , 4 \alpha), \ \rightarrow \ t_2, \\
(-2 , -4 \alpha), \ \rightarrow \ t_3, \\
(-2 , 0), \ \rightarrow \ t_4,
\end{cases}
\end{eqnarray}
the corresponding probabilities for each set of values are indicated in front of them.

Similarly, for $1<\alpha<2$ we obtain the possible values for $(a-b,a-d)$
\begin{equation}
\begin{cases}
(2,4) &\to u_1, \\
(-2,-4) &\to u_2, \\
(-2,2-2\alpha) &\to u_3, \\
(2,-4+4\alpha) &\to u_4.
\end{cases}\label{probs12}
\end{equation}
For $2<\alpha<3$ the possible values for $(a-b,a-d)$ are
\begin{equation}
\begin{cases}
(2,4) &\to v_1, \\
(-2,-4) &\to v_2, \\
(-2,2-2\alpha) &\to v_3. \\
\end{cases}\label{probs23}
\end{equation}
Finally, for $\alpha>3$ we obtain
\begin{equation}
(a-b,a-d)=\begin{cases}
(2,4) &\to w_1, \\
(-2,-4) &\to w_2.  \\
\end{cases}\label{probs3}
\end{equation}
In the next section we will derive the relations between probabilities and their solutions.

\section{Invariant measure}\label{Inv-measure}
For a long sequence of matrices, the probabilities of occurrence of different values of the exponents become stationary. According to the rules of evolution, presented in the Appendix A, for $\alpha<1$ the probabilities satisfy the following set of equations
\begin{eqnarray}
&&p_0 = r_0, \nonumber\\
&&p_1 = (1-x) (r_1 + q_1),  \nonumber\\
&&p_2 = (1-x) (p_0 + q_2 + t_1),  \nonumber\\
&&p_4 = (1-x) (p_2 + q_4 + t_2),  \nonumber\\
&&p_i = (1-x) (p_{i-2} + q_i), \   4< i < n, \label{eqpi}\\
&&p_i = (1-x) p_{i-2},  \  n \leq i < 2n, \label{eqpi2}\\
&&p_{2n} = (1-x) (p_{2n-2} + p_{2n-1} + p_{2n} ),  \nonumber\\
&&q_i = x r_i, \   0 \leq i \leq n-1, \label{eqqi}\\
&&r_i = (1-x) (r_{i+2} + s_i), \   0< i < n-2, \label{eqri}\\
&&r_{n-1} = (1-x) s_{n-1},  \nonumber\\
&&r_{n-2} = (1-x) s_{n-2},  \nonumber\\
&&s_0 = x r_0,  \nonumber\\
&&s_i = x p_i, \  0 < i < n-1, \label{eqsi}\\
&&s_{n-1} = x \sum_{i=n-1}^{2 n} p_i,  \nonumber\\
&&t_1 = (1-x) t_3,   \nonumber\\
&&t_2 = (1-x) t_4,   \nonumber\\
&&t_3 = x (t_1 + t_2),  \nonumber\\
&&t_4 = x \left(\sum_{i=1}^{n-1} s_i + \sum_{i=0}^{n-1} q_i + t_3 + t_4 \right)\nonumber.
\end{eqnarray}
This set of equations allows the following closed solution
\begin{eqnarray}
p_{i}&=&
\begin{cases}
c_1 \lambda_1^i + c_2 \lambda_2^i, \ 0 < \text{even} \ i < n,  \\
c_3 \lambda_1^i + c_4 \lambda_2^i, \ 0 < \text{odd} \ i < n, \\
c_5 \lambda_3^i, \  n \leq \text{even} \ i < 2n, \\
c_6  \lambda_3^i,  \ n \leq \text{odd} \ i < 2n,
\end{cases} \nonumber\\
r_i&=&
\begin{cases}
c_7 \lambda_1^i + c_8 \lambda_2^i, \ 0 < \text{even} \ i < n, \\
c_9 \lambda_1^i + c_{10} \lambda_2^i,  \ 0 < \text{odd} \ i < n,
\end{cases} \nonumber\\
q_i&=&xr_i, \nonumber\\
s_i&=&xp_i, \nonumber\\
t_1 &=& \frac{x^3 (1-x)^2}{1-x+x^2}, \nonumber\\
t_2 &=& x^2 (1-x), \nonumber\\
t_3 &=&  \frac{x^3 (1-x)}{1-x+x^2}, \nonumber\\
t_4 &=& x^2,
\label{probs}
\end{eqnarray}
where
\begin{eqnarray*}
\lambda_1 &=& \frac{1}{\sqrt{1-x}}\Big( 1 - x +
x^3 - \frac12 x^4 \\
&&- \frac12 x^{\frac{3}{2}} \sqrt{8 - 12 x + 4 x^2 + 4 x^3 - 4 x^4 + x^5} \Big)^{\frac{1}{2}}, \\
\lambda_2 &=& \frac{1}{\sqrt{1-x}}\Big( 1 - x +
x^3 - \frac12 x^4 \\
&&+ \frac12 x^{\frac{3}{2}} \sqrt{8 - 12 x + 4 x^2 + 4 x^3 - 4 x^4 + x^5} \Big)^{\frac{1}{2}},\\
\lambda_3 &=& \sqrt{1-x}.
\end{eqnarray*}
$\lambda_1$ and $\lambda_2$ are the roots of the characteristic equation
\begin{equation}\label{char1}
\left( (1-x) - \lambda^2  \right) \left( (1-x) \lambda^2 -1 \right) - x^2 (1-x)^2 \lambda^2 =0,
\end{equation}
and $\lambda_3$ is the root of
\begin{equation}\label{char2}
\lambda^2 =1-x.
\end{equation}

The Eq. (\ref{char1}) is obtained by replacing $q_i$ from Eq. (\ref{eqqi}) in Eq. (\ref{eqpi}) and $s_i$ from Eq. (\ref{eqsi}) in Eq. (\ref{eqri}) and then using the ansatz $p_i\propto \lambda^i$ and $r_i\propto \lambda^i$ and setting the determinant of the coefficients of the resulting equations, equal to zero. Equation (\ref{char2}) is obtained from Eq. (\ref{eqpi2}) using the ansatz $p_i\propto \lambda^i$. Since the even and odd indices are separated in Eqs. (\ref{probs}), the roots with the negative signs will give the same solution.

The coefficients $c_1,\dots, c_{10}$  should be determined in terms  of  $x$ and $n$. Substituting the above solution in the corresponding equations we obtain a set of equations for $c_i$s which are given in Appendix B.

For other ranges of $\alpha$ the equations and their solutions are as follows. For $1<\alpha<2$,
\begin{eqnarray}\label{probsol12}
\begin{cases}
u_1 = (1-x) (u_1 + u_3 + u_4), \\
u_2 = x (u_1 + u_4), \\
u_3 = x (u_2 + u_3), \\
u_4 = (1-x) u_2,
\end{cases} \notag\\  \Rightarrow
\begin{cases}
u_1 = (1-x)(1-x+x^2), \\
u_2 = x (1-x), \\
u_3 = x^2, \\
u_4 = x (1-x)^2.
\end{cases}
\end{eqnarray}
For $2<\alpha<3$,
\begin{eqnarray}\label{probsol23}
\begin{cases}
v_1 = (1-x) (v_1 + v_2 + v_3), \\
v_2 = x v_1, \\
v_3 = x (v_2 + v_3),\\
\end{cases} \notag\Rightarrow
\begin{cases}
v_1 = (1-x), \\
v_2 = x (1-x), \\
v_3 = x^2.\\
\end{cases}\notag \\
\end{eqnarray}
For $\alpha>3$,
\begin{eqnarray}\label{probsol3}
\begin{cases}
w_1 = (1-x) (w_1 + w_2), \\
w_2 = x (w_1 + w_2), \\
\end{cases} \notag\Rightarrow
\begin{cases}
w_1 = 1-x, \\
w_2 = x . \\
\end{cases}\notag \\
\end{eqnarray}

\section{Ground state energy and magnetization}
Now the ground state energy can be calculated via \cite{derrida1978simple}
\begin{eqnarray}  \notag
E/J &=& -\frac12 \lim_{L \to \infty} \langle a \rangle_L / L =  -\frac12 \lim_{L \to \infty} ( \langle a \rangle_{L+1} - \langle a \rangle_L ), \\
&=&  -\frac12 \lim_{L \to \infty}  \langle a_{L+1}-a_L \rangle,
\label{energy-a}
\end{eqnarray}
and differentiating energy with respect to the field gives magnetization $m=-\partial E/\partial h=-\partial (E/J)/\partial \alpha$. The factor $\frac12$ is included to make the energy per spin.

The infinite length limit in Eq. (\ref{energy-a}) can be replaced with an average with respect to stationary solution for probabilities given in Eqs. (\ref{probs12}),(\ref{probs23}),(\ref{probs3}) and (\ref{probs}). For $\alpha<1$, using the equations in Appendix A we need to calculate the average of $a_{L+1}-a_L=A-a$. We should sum up different values of $A-a$ multiplied with corresponding probabilities. For instance in the first set of equations (with probability $1-x$) we can see that $A-a=3+2\alpha$ in all cases. Therefore from the first set we only get the contribution $(3 + 2 \alpha)(1-x)$. Including the contributions from the second set (with probability $x$) we obtain
\begin{eqnarray}  \notag
E/J &=& -\frac{1}{2}(3 + 2 \alpha) (1-x) -\frac{1}{2} x \Big[\sum_{i=n-1}^{2n} (-2 \alpha n+1 ) p_i \\ \notag
&& + \sum_{i=0}^{n-2}  (-2 \alpha (i+1) + 1 )p_i + \sum_{i=1}^{n-1}  (2 \alpha (i-1) + 1 )r_i  \\ \notag
&& + \sum_{i=0}^{n-1}  (2 \alpha (i+1) + 1 )s_i + \sum_{i=1}^{n-1}  (2 \alpha + 1 )q_i  \\
&& + (1-2 \alpha)(t_1 + t_2) + 3(t_3 + t_4) \Big].  \label{energy-p}
\end{eqnarray}
For $1<\alpha<3$, using Eqs. (\ref{probs12}), (\ref{probs23}), (\ref{probsol12}) and (\ref{probsol23})
\begin{equation}
E/J = -\frac{1}{2} (3(2 x^2 -2 x +1)+2 \alpha (1-x^2)),\
\end{equation}
and then by differentiating with respect to $\alpha$ the magnetization is
\begin{equation}
m = 1-x^2.
\end{equation}
For $\alpha>3$, using Eqs. (\ref{probs3}) and (\ref{probsol3}), the energy is
\begin{equation}
E/J = -\frac{1}{2} ((3+2\alpha)(1-x)+(-3+2\alpha)x),
\label{E3}
\end{equation}
then by differentiating Eq. \eqref{E3} with respect to $\alpha$, one finds that the magnetization is saturated ($m=1$) as expected.

\section{Numerical calculations}
In order to confirm our analytical results we repeat the calculations using a numerical method. We use the iterative approach which is applied to the quasi-one dimensional random Ising model and is based on the following set of coupled equations \cite{sepehrinia2018ground}
\begin{eqnarray}\label{energy}
E^{\mu}_{l}=\min_{\nu}\{E^{\nu}_{l-1}-\sum_{j} J^h_{l-1,j}\sigma^{\nu}_{l-1,j} \sigma^{\mu}_{l,j}\} \hspace{2cm} \nonumber \\ - \sum_{j} J^v_{lj}\sigma^{\mu}_{l,j} \sigma^{\mu}_{l,j+1} -
  h \sum_j  \sigma^{\mu}_{l,j},\nonumber \\
\end{eqnarray}
where $E^{\mu}_{l}$s are the ground state energies of the strip of length $l$ for a given configuration, $\mu$, of spins in the last column.

\section{Results and discussion}\label{reaults}
\begin{figure}[h]
  		\begin{center}
  				\includegraphics[scale=0.8]{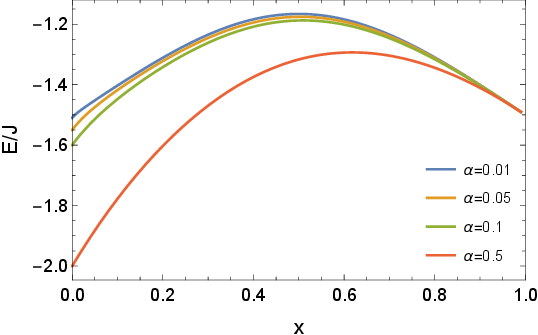}
  		\end{center}
  		\caption{Ground state energy per spin as a function of the concentration for different values of magnetic field ($\alpha=h/J$).}
  		\label{E-x}
\end{figure}

The ground state energy per spin as a function of the concentration of antiferromagnetic bonds for different values of the uniform magnetic field is shown in Fig. \ref{E-x}. For zero magnetic field ($\alpha \rightarrow 0$) the curve is symmetric \cite{kadowaki1996exact} under the exchange of $x$ and $(1-x)$ , and has a maximum at $x=\frac{1}{2}$. The minimum energy corresponds to $x=0$ and $x=1$ because there is no frustration in these cases. By applying a magnetic field, this symmetry breaks. At $x=1$ the magnetic field does not split the energy because all bonds are antiferromagnetic and magnetization is zero for small magnetic fields, therefore the energy does not change by increasing the magnetic field up to a threshold value. In the presence of the magnetic field, the minimum energy corresponds to the ferromagnetic ladder ($x=0$) and the concentration with maximum energy shifts to higher values of $x$.

\begin{figure}[h]
  		\begin{center}
  				\includegraphics[scale=0.8]{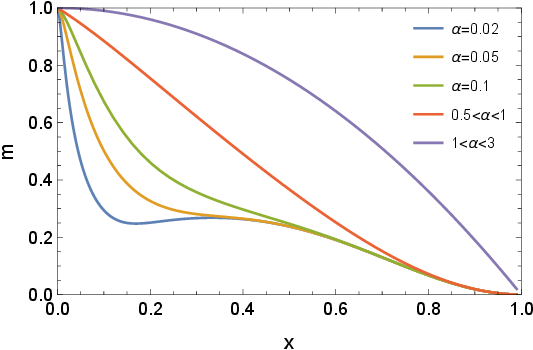}
  		\end{center}
  		\caption{Magnetization per spin as a function of the concentration for different values of magnetic field ($\alpha=h/J$).}
  		\label{m-x}
\end{figure}

In Fig. \ref{m-x} we have plotted the magnetization as a function of the concentration of antiferromagnetic bonds. At small magnetic fields, magnetization has a nonmonotonic dependence on concentration and exhibits a local minimum, but at higher magnetic fields the behavior is different and the minimum disappears. Magnetization tends to zero by increasing the concentration of antiferromagnetic bonds due to antiferromagnetic order at higher values of $x$. It is interesting to compare Fig. \ref{m-x} with the same result for the single chain (Fig. 1 of Ref. \cite{derrida1978simple}). We see that nonmonotonic behavior is peculiar to the ladder. In order to see whether the nonmonotonicity persists by changing the randomness or not, we calculated the magnetization with all eight matrices and it turns out that it disappears at least for the magnetic field of the order $h\sim 0.01$.
\begin{figure}[h]
  		\begin{center}
  				\includegraphics[scale=0.8]{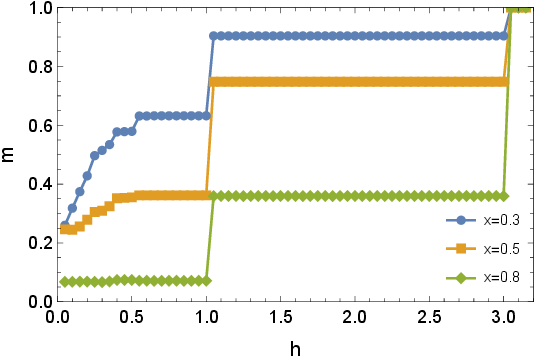}
  		\end{center}
  		\caption{Numerical results for magnetization as a function of the magnetic field for different values of concentration.}
  		\label{m-h}
\end{figure}
\begin{figure}[b]
  		\begin{center}
  				\includegraphics[scale=0.8]{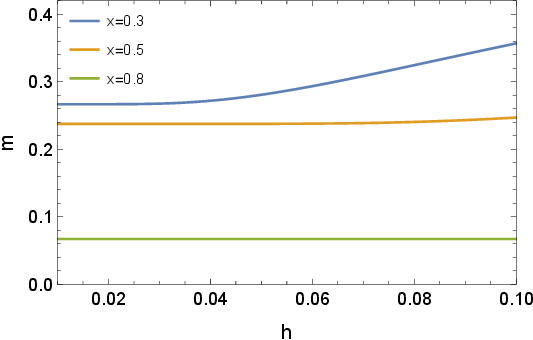}
  		\end{center}
  		\caption{Magnetization as a function of the magnetic field for different values of concentration.}
  		\label{m-h-0}
\end{figure}

Figure \ref{m-h} shows the numerical results for magnetization as a function of the magnetic field for different values of the concentration. The step-like behavior is observed as in the single chain, although the jumps occur at different sets of magnetic fields. However, the magnetization behaves differently in zero magnetic field limit as it tends to a nonzero value depending on the concentration. We have plotted the analytical result for small magnetic fields in Fig. \ref{m-h-0}. It can be seen that the limiting values for $h\rightarrow 0$ agree in Fig. \ref{m-h} and Fig. \ref{m-h-0}. The precise functional form of magnetization in the zero field limit can be revealed using our analytical result. As we explained in section \ref{Inv-measure} we obtain the probabilities in terms of $x$ and $n$. Then we insert them in Eq. (\ref{energy-a}) and by doing the summations and keeping the leading terms as $n\rightarrow \infty$, we find that the energy has an essential singularity known as Griffiths singularity
\begin{equation}
E \simeq E_0 + m_0 h + E_1 e^{-h_0/h}, \ h\rightarrow 0,
\label{GriffE}
\end{equation}
from which we obtain
\begin{equation}
m \simeq m_0 +  m_1 \frac{e^{-h_0/h}}{h^2}, \ h\rightarrow 0,
\label{Griffm}
\end{equation}
where $m_1=E_1 h_0$ and $m_0$ is the zero field magnetization. Similar singularity is known to exist in diluted Ising ferromagnet \cite{wortis1974griffiths,imry1977griffiths} and higher dimensional random temperature Ginzburg-Landau Hamiltonian \cite{dotsenko2006griffiths}. The existence of the Griffiths phase in the spin glass is also suggested by studying the dynamics \cite{randeria1985low} and distribution of Lee-Yang zeros in equilibrium \cite{matsuda2008distribution}. It is argued in the Refs. \cite{randeria1985low,matsuda2008distribution} that a common physics underlies the singularity as in the diluted ferromagnets, namely, the large-sized connected clusters. Although the connected clusters are not trivially defined in the spin glass. We believe the singularity that we have found here has the same cause. In our case, the magnetic field dependence of the magnetization is governed by the large ferromagnetic clusters that occur with an exponentially small probability. Our model allows an estimation of this effect as follows. The probability of the ferromagnetic cluster of length $\ell$ is of order $(1-x)^{\ell}$. Such a cluster will flip when the magnetic field is such that $\ell h\sim J$ so the change in magnetization will be proportional to $(1-x)^{J/h}$. This is, of course, a crude estimate that only shows how the essential singularity arises, but our result in Eq. (\ref{Griffm}) shows the true nature of the singularity.

\begin{figure}[t]
  		\begin{center}
  				\includegraphics[scale=0.8]{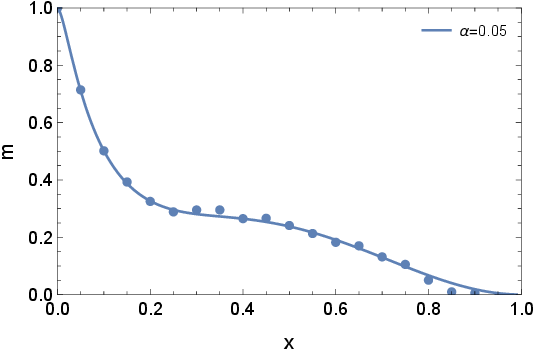}
  		\end{center}
  		\caption{Magnetization as a function of the concentration for $\alpha=0.05$. The solid curve is the analytical result and circles show the numerical result.}
  		\label{m-x_nu-an}
\end{figure}

For a comparison of analytical results with the results of numerical computations we have plotted magnetization as a function of concentration in Fig. \ref{m-x_nu-an}.

In conclusion, we presented analytical results on the zero temperature properties of the restricted $\pm J$ Ising spin glass ladder in the presence of the uniform magnetic field. We found that, in contrast to the single chain model, magnetization of the frustrated ladder tends to nonzero value by decreasing the magnetic field. Another notable feature of magnetization of the ladder is that depending on the randomness and magnetic field, the magnetization can be a nonmonotonic function of concentration. We also found that, thermodynamic quantities are not analytic functions of magnetic field as $h \rightarrow 0$ which is known as Griffiths singularity. We derived the explicit functional form of the singularity. The accuracy of analytical results was demonstrated by performing numerical calculations.

As we mentioned earlier, we expect to see the similar behavior in the original $\pm J$ model because of the following argument. The singularity results from certain large sequences of matrices in the ladder. Adding other matrices in the ladder does not eliminate those sequences. They can still occur however with different probability. Therefore, we expect the singularity to be generated by those sequences although with different strength. The other feature of the magnetization is the zero field limit of it $m_0$. We expect this also remains nonzero in the case of the original 8-matrix model because there will be a finite concentration of frustrated plaquettes therefore the ground state will be degenerate and a small magnetic field induces a finite magnetization. We support this predictions with numerical calculation of magnetization of the 8-matrix model which shows the same features that we obtained in two matrix model.

Our solution opens a way to study the models with more realistic disorders and higher widths and also can be used to study the accuracy of optimization algorithms of finding the spin glass ground state.

\section{acknowledgement}
We would like to acknowledge financial support from the research council of University of Tehran. This work is based upon research funded by Iran National Science Foundation (INSF) under project No.4005950.
\appendix
\addcontentsline{toc}{section}{Appendices}
\section{}
In this appendix we derive the equations of evolution for the exponents of the two matrix model with $M_1$ and $M_2$, given in Eq. (\ref{M1M2}), with probabilities $1-x$ and $x$ respectively and magnetic fields $\alpha=1/n$ for $n=2,3,\dots$ ($\alpha=h/J$).

By keeping the leading term in the elements of the first row of $\mathcal{M}_L $ after being multiplied by one of the matrices, using Eq. (\ref{TM}) and Eq. (\ref{Matrow}) the new exponents can be expressed in terms of the old ones as follows. After being multiplied by $M_1$
\begin{eqnarray*}
A &=& \max(a+3+2\alpha ; b-1 ; c-1 ; d-1-2\alpha), \\
B &=& \max(a+1+2\alpha ; b+1 ; c-3 ; d+1-2\alpha), \\
C &=& \max(a+1+2\alpha ; b-3 ; c+1 ; d+1-2\alpha), \\
D &=& \max(a-1+2\alpha ; b-1 ; c-1 ; d+3-2\alpha),
\end{eqnarray*}
and after being multiplied by $M_2$
\begin{eqnarray*}
A &=& \max(a-3+2\alpha ; b+1 ; c+1 ; d+1-2\alpha), \\
B &=& \max(a-1+2\alpha ; b-1 ; c+3 ; d-1-2\alpha), \\
C &=& \max(a-1+2\alpha ; b+3 ; c-1 ; d-1-2\alpha), \\
D &=& \max(a+1+2\alpha ; b+1 ; c+1 ; d-3-2\alpha).
\end{eqnarray*}
Then using $b=c$ (as we explained in the text) we find with probability $(1-x)$

\begin{eqnarray}
\text{if} &\begin{cases}\label{M1A1}
  -2 \alpha \leq a-b \leq 2 \\
 4-4 \alpha \leq a-d \leq 4 \\
 \end{cases} &\Rightarrow
\begin{cases}
A = a+3+2\alpha \\
B = a+1+2\alpha \\
D = a-1+2\alpha
\end{cases} \\
\text{if} &\begin{cases}
  -2 \alpha \leq a-b \leq 2 \\
 -4 \alpha \leq a-d \leq 4 -4\alpha \\
 \end{cases} &\Rightarrow
\begin{cases}
A = a+3+2\alpha \\
B = a+1+2\alpha \\
D = d+3-2\alpha
\end{cases} \\
\text{if} &\begin{cases}
  -2 \alpha \leq a-b \leq  2 \\
 -4  \leq a-d \leq  -4\alpha \\
 \end{cases} &\Rightarrow
\begin{cases}\label{M1A3}
A = a+3+2\alpha \\
B = d+1-2\alpha \\
D = d+3-2\alpha
\end{cases} \\
\text{if} &\begin{cases}
  -2 \leq a-b \leq -2 \alpha  \\
4 - 4\alpha \leq a-d \leq 4 \\
 \end{cases} &\Rightarrow
\begin{cases}
A = a+3+2\alpha \\
B = b+1 \\
D = b-1
\end{cases} \\
\text{if} &\begin{cases}
  -2 \leq a-b \leq -2 \alpha  \\
- 4\alpha \leq a-d \leq 4 - 4\alpha  \\
 \end{cases} &\Rightarrow
\begin{cases}
A = a+3+2\alpha \\
B = b+1 \\
D = d+3-2\alpha
\end{cases} \\
\text{if} &\begin{cases}
  -2 \leq a-b \leq -2 \alpha  \\
- 4 \leq a-d \leq - 4\alpha  \\
- 2 \alpha \leq b-d \leq 2
 \end{cases} &\Rightarrow
\begin{cases}
A = a+3+2\alpha \\
B = b+1 \\
D = d+3-2\alpha
\end{cases} \\
\text{if} &\begin{cases}
  -2 \leq a-b \leq -2 \alpha  \\
- 4 \leq a-d \leq - 4\alpha  \\
- 2 \leq b-d \leq - 2 \alpha
 \end{cases} &\Rightarrow
\begin{cases}
A = a+3+2\alpha \\
B = d+1- 2 \alpha  \\
D = d+3-2\alpha
\end{cases}
\end{eqnarray}

and with probability $x$ we have
\begin{eqnarray}
\text{if} &\begin{cases}
  -2 \alpha \leq a-b \leq 2 \\
 4-4 \alpha \leq a-d \leq 4 \\
 \end{cases} &\Rightarrow
\begin{cases}
A = b+1 \\
B = b+3 \\
D = a+1+2\alpha
\end{cases} \\
\text{if} &\begin{cases}
  -2 \alpha \leq a-b \leq 2 \\
 -4 \alpha \leq a-d \leq 4 -4\alpha \\
 - 2 \alpha \leq b-d \leq 2
 \end{cases} &\Rightarrow
\begin{cases}
A = b+1 \\
B = b+3 \\
D = a+1+2\alpha
\end{cases} \\
\text{if} &\begin{cases}
  -2 \alpha \leq a-b \leq 2 \\
 -4 \alpha \leq a-d \leq 4 -4\alpha \\
 - 2  \leq b-d \leq - 2 \alpha
 \end{cases} &\Rightarrow
\begin{cases}
A = d+1-2\alpha  \\
B = b+3 \\
D = a+1+2\alpha
\end{cases} \\
\text{if} &\begin{cases}
  -2 \alpha \leq a-b \leq 2 \\
 -4  \leq a-d \leq -4\alpha \\
 \end{cases} &\Rightarrow
\begin{cases}
A = d+1-2\alpha \\
B = b+3 \\
D = a+1+2\alpha
\end{cases} \\
\text{if} &\begin{cases}
  -2 \leq a-b \leq -2 \alpha  \\
- 4\alpha \leq a-d \leq 4 \\
 \end{cases} &\Rightarrow
\begin{cases}
A = b+1 \\
B = b+3 \\
D = b+1
\end{cases} \\
\text{if} &\begin{cases}
  -2 \leq a-b \leq -2 \alpha  \\
- 4 \leq a-d \leq - 4\alpha  \\
- 2 \alpha \leq b-d \leq 2
 \end{cases} &\Rightarrow
\begin{cases}
A = b+1 \\
B = b+3 \\
D = b+1
\end{cases} \\
\text{if} &\begin{cases}
  -2 \leq a-b \leq -2 \alpha  \\
- 4 \leq a-d \leq - 4\alpha  \\
- 2 \leq b-d \leq - 2 \alpha
 \end{cases} &\Rightarrow
\begin{cases}
A = d+1-2\alpha \\
B = b+3  \\
D = b+1
\end{cases}
\end{eqnarray}

\section{}

In this appendix, we shall give the equations for coefficients $c_1,\dots , c_{10}$.
\begin{equation} \label{eq-app1}
c_1 = \frac{x(1-x)\lambda_1^2}{\lambda_1^2-(1-x)} c_7,
\end{equation}
\begin{equation} \label{eq-app2}
c_2 = \frac{x(1-x)\lambda_2^2}{\lambda_2^2-(1-x)} c_8.
\end{equation}
\begin{equation}
c_3 = \frac{x(1-x)\lambda_1^2}{\lambda_1^2-(1-x)} c_9,
\end{equation}
\begin{equation}
c_4 = \frac{x(1-x)\lambda_2^2}{\lambda_2^2-(1-x)} c_{10}.
\end{equation}
\begin{equation}
c_5=\begin{cases}
\left( \frac{\lambda_1}{\lambda_3} \right)^{n-2} c_1 + \left( \frac{\lambda_2}{\lambda_3} \right)^{n-2} c_2, \ \text{even} \ n, \\
\left( \frac{\lambda_1}{\lambda_3} \right)^{n-1} c_1 + \left( \frac{\lambda_2}{\lambda_3} \right)^{n-1} c_2, \ \text{odd} \ n
\end{cases}
\end{equation}
\begin{equation}
c_6=\begin{cases}
\left( \frac{\lambda_1}{\lambda_3} \right)^{n-1} c_3 + \left( \frac{\lambda_2}{\lambda_3} \right)^{n-1} c_4, \ \text{even} \ n, \\
\left( \frac{\lambda_1}{\lambda_3} \right)^{n-2} c_3 + \left( \frac{\lambda_2}{\lambda_3} \right)^{n-2} c_4, \ \text{odd} \ n
\end{cases}
\end{equation}
For even $n$,
\begin{equation}
c_1 = \left( \frac{\lambda_2}{\lambda_1} \right)^{n-4} \left( \frac{(1-x) - \lambda_2^2 (1-x^2 (1-x)^2)}{(1-x^2 (1-x)^2)\lambda_1^2 - (1-x)} \right) c_2,
\end{equation}
For odd $n$,
\begin{equation}
c_3 = \left( \frac{\lambda_2}{\lambda_1} \right)^{n-4} \left( \frac{(1-x) - \lambda_2^2 (1-x^2 (1-x)^2)}{(1-x^2 (1-x)^2)\lambda_1^2 - (1-x)} \right) c_4,
\end{equation}
\begin{equation}
c_3 \lambda_1 + c_4 \lambda_2 = (1-x^2) \left( c_9 \lambda_1 + c_{10} \lambda_2 \right).
\end{equation}
For even $n$,
\begin{eqnarray}
c_9 \lambda_1^{n-1} + c_{10} \lambda_2^{n-1} = & \notag \\
x (1-x)& \left(  c_5 \sum_{k=\frac{n}{2}}^{n-2} \lambda_3^{2 k} + c_6 \sum_{k=\frac{n}{2}}^{n-1} \lambda_3^{2 k-1} \right) \notag \\
+ (1-x) & \left(  c_5 \lambda_3^{ 2n -2}  + c_6 \lambda_3^{ 2n -1}  \right). \notag \\
\end{eqnarray}
For odd $n$,
\begin{eqnarray}
c_7 \lambda_1^{n-1} + c_{8} \lambda_2^{n-1} = & \notag \\
x (1-x)& \left(  c_5 \sum_{k=\frac{n-1}{2}}^{n-2} \lambda_3^{2 k} + c_6 \sum_{k=\frac{n+1}{2}}^{n-1} \lambda_3^{2 k-1} \right) \notag \\
+ (1-x)& \left(  c_5 \lambda_3^{ 2n -2}  + c_6 \lambda_3^{ 2n -1}  \right), \notag \\
\end{eqnarray}
Normalization of the probabilities,
\begin{eqnarray}
\sum_{i=1}^{2 n} p_i + \sum_{i=0}^{n-1} (q_i+r_i) + \sum_{i=1}^{n-1} s_i +t_1 +t_2 + t_3 + t_4 = 1, \notag \\
\end{eqnarray}
gives, for even $n$,
\begin{eqnarray}
&  \sum_{k=2}^{\frac{n}{2}-1} (\lambda_1^{2 k} c_1 +  \lambda_2^{2 k} c_2 +   \lambda_1^{2 k-1} c_3) \notag \\
&+c_4 \sum_{k=2}^{\frac{n}{2}-1} \lambda_2^{2 k-1}+c_5  \sum_{k=\frac{n}{2}}^{n-2} \lambda_3^{2 k} + c_6 \sum_{k=\frac{n}{2}}^{n-1}  \lambda_3^{2 k-1}    \notag \\
&+ \frac{\lambda_3^{2 n-2}}{x} (c_5+ \lambda_3 c_6) + \left( \frac{2-x}{(1-x)^2}  -x^2 +x\right) \left( c_1 \lambda_1^4 + c_2 \lambda_2^4 \right) \notag \\
&+  \sum_{k=2}^{\frac{n}{2}-1} (\lambda_1^{2 k} c_7 + \lambda_2^{2 k} c_8) +   \sum_{k=2}^{\frac{n}{2}} (\lambda_1^{2 k-1}c_9 +  \lambda_2^{2 k-1}c_{10}) \notag \\
&+ \left( \frac{x}{(x-1)} + (1-x)^2 (1-x^2) - x  \right) \left( c_7 \lambda_1^4 + c_8 \lambda_2^4 \right) \notag \\
&+ \left( \frac{(2-x^2)(1-x)}{1-x(1-x)(1-x^2)}  \right) \left( c_9 \lambda_1^3 + c_{10} \lambda_2^3 \right) \notag \\
& - \left(1+ x(1-x) + \frac{1}{(1-x)} + x^2 (1-x) \right) x^2 (1-x) \notag  \\
&- \frac{x^3(1-x)^2}{1-x+x^2} = \frac{(1-x)^2}{1-x+x^2}, \notag \\
\end{eqnarray}
and for odd $n$,
\begin{eqnarray}
&   \sum_{k=2}^{\frac{n-3}{2}} (\lambda_1^{2 k} c_1 +  \lambda_2^{2 k} c_2) +   \sum_{k=2}^{\frac{n-1}{2}} (\lambda_1^{2 k-1}c_3 + \lambda_2^{2 k-1} c_4) \notag \\
&+c_5  \sum_{k=\frac{n-1}{2}}^{n-2} \lambda_3^{2 k} + c_6 \sum_{k=\frac{n+1}{2}}^{n-1}  \lambda_3^{2 k-1} + \frac{1}{x} c_5 \lambda_3^{2 n-2}  \notag \\
&+\frac{1}{x} c_6 \lambda_3^{2 n-1}+ \left( \frac{1}{(1-x)^2} +\frac{1}{(1-x)} -x^2 +x\right) \left( c_1 \lambda_1^4 + c_2 \lambda_2^4 \right) \notag \\
&+  \sum_{k=2}^{\frac{n-1}{2}} (\lambda_1^{2 k} c_7 +  \lambda_2^{2 k} c_8 +   \lambda_1^{2 k-1} c_9  +  \lambda_2^{2 k-1} c_{10}) \notag \\
&+ \left( \frac{x}{(x-1)} + (1-x)^2 (1-x^2) - x  \right) \left( c_7 \lambda_1^4 + c_8 \lambda_2^4 \right) \notag \\
&+ \left( \frac{(2-x^2)(1-x)}{1-x(1-x)(1-x^2)}  \right) \left( c_9 \lambda_1^3 + c_{10} \lambda_2^3 \right) \notag \\
& - \left(1+ x(1-x) + \frac{1}{(1-x)} + x^2 (1-x) \right) x^2 (1-x) \notag  \\
&- \frac{x^3(1-x)^2}{1-x+x^2} = \frac{(1-x)^2}{1-x+x^2}. \notag  \\
\end{eqnarray}

\bibliography{ref_Izadi2}
\bibliographystyle{apsrev}

\end{document}